\documentclass[prd,aps,preprint,epsfig,floats,showaps]{revtex4}
\usepackage{amsmath}
\usepackage{amsfonts}
\usepackage{graphicx}
\usepackage{epsfig}
\begin{document}
\newcommand{\vect}[1]{\overrightarrow{#1}}
\newcommand{\smbox}[1]{\mbox{\scriptsize #1}}
\newcommand{\slh}[1]{\displaystyle{\not}#1}
\newcommand{\gL}{g_{\mbox{\tiny{L}}}}
\newcommand{\gR}{g_{\mbox{\tiny{R}}}}
\newcommand{\gY}{g_{\mbox{\tiny{Y}}}}
\newcommand{\gBL}{g_{\mbox{\tiny{BL}}}}
\newcommand{\gdenomA}{\sqrt{\gL^2\gR^2+\gBL^2\gL^2+\gBL^2\gR^2}}
\newcommand{\gdenomB}{\sqrt{\gBL^2+\gR^2}}
\newcommand{\Psibar}{\overline{\Psi}}
\newcommand{\Bvec}{B^{\smbox{(11)}}_{Y,\mu}}
\newcommand{\Bsca}{B^{\smbox{(11)}}_{Y,(-)}}
\preprint{\vbox{ \hbox{UMD-PP-06-004} }}
\title{\Large\bf Dark Matter in Universal Extra Dimension Models:
$B_Y^{KK}$ vs. $\nu_{R}^{KK}$}
\author{\bf Ken Hsieh, R.N. Mohapatra and Salah Nasri }

\affiliation{ Department of Physics, University of Maryland, College
Park, MD 20742, USA}

\date{July, 2006}

\begin{abstract}
We show that in a class of universal extra dimension models (UED),
which solves both the neutrino mass and proton decay problem, an
admixture of KK photon and KK right handed neutrinos can provide the
required amount of cold dark matter (CDM). This model has two
parameters $R^{-1}$ and $M_{Z'}$ ($R$ is the radius of the extra
space dimensions and $Z'$ the extra neutral gauge boson of the
model). Using the value of the relic CDM density, combined with the
results from the cryogenic searches for CDM, we obtain upper limits
on $R^{-1}$ of about $400-650$ GeV and $M_{Z'}\leq 1.5$ TeV, both
being accessible to LHC. In some regions of the parameter space, the
dark matter-nucleon scattering cross section can be as high as of
$10^{-44}$ cm$^2$, which can be probed by the next round of dark
matter search experiments.
\end{abstract}
\pacs{12.60.-i, 11.10.Kk, 95.35.+d} \maketitle
\section{Introduction}
The existence of dark matter is now a well established fact. The
nature of the dark matter particle is however still a mystery. Its
discovery is going to be a major breakthrough in the study of
physics beyond the standard model of both particle physics and
cosmology. One candidate that has rightly received a great deal of
attention is the lightest supersymmetric particle, the neutralino,
since there are many reasons to think that physics may be
supersymmetric around the TeV scale. Furthermore, the Large Hadron
Collider machine at CERN, which is scheduled to start operating in
mid-2007,  will experimentally explore physics at the TeV energy
scale making it possible to have a detailed understanding of dark
matter related physics.

Another class of models which leads to a very different kind of TeV
scale physics and will also be explored at LHC is one where there
exist extra space dimensions with sizes of order of an inverse TeV.
In particular there is a class of extra dimension models known as
universal extra dimension models (UED) where all standard model
particles live in either five (or six) dimensional space time of
which one (or two) is (are) compactified with radius $R^{-1}\leq $
TeV\cite{acd}. A recently discussed cosmologically interesting point
about the UED models \cite{tait,feng} is that the lightest
Kaluza-Klein particles of these models being stable can serve as
viable dark matter candidates. This result is nontrivial due to the
fact that the dark matter relic abundance is determined by the
interactions in the theory which are predetermined by standard
model. It turns out that the first KK mode of the hypercharge boson
is the dark matter candidate provided the inverse size of the extra
dimension is less than a TeV.

A generic phenomenological problem with 5D UED models based on the
standard model gauge group is that they can lead to rapid proton
decay as well as unsuppressed neutrino masses. A way to cure the
rapid proton decay problem is to consider six dimensions\cite{yee}
where the extra space dimensions lead to a new $U(1)$ global
symmetry that suppresses the strength of all baryon number
nonconserving operators. On the other hand both the neutrino mass
and the proton decay problem can be solved simultaneously if we
extend the gauge group of the six dimensional model to
$SU(2)_L\times SU(2)_R\times U(1)_{B- L}$\cite{abdel}. With
appropriate orbifolding, a neutrino mass comes out to be of the
desired order due to a combination two factors: the existence of
$B-L$ gauge symmetry and orbifolding which keeps the left-handed
singlet neutrino as a zero mode forbid the lower dimensional
operators that could give neutrino mass. Another advantage of the 6D
models over the 5D ones is that cancellation of gravitational
anomaly automatically leads to the existence of the right handed
neutrinos needed for generating neutrino masses.

In this paper, we point out that the 6D UED models with an extended
gauge group of ref. \cite{abdel} provide a two-component picture of
dark matter consisting of a KK right handed neutrino and a KK
hypercharge boson. We do a detailed calculation of the relic
abundance of both the $\nu_{R,KK}$ and the $B_Y^{KK}$ as well as the
cross section for scattering of the dark matter in the cryogenic
detectors in these models. The two main results of this calculation
are that: (i) present experimental limits \cite{cdms} on the
DM-nucleon cross section and the value of the relic
density\cite{wmap} imply very stringent limits on the the two
fundamental parameters of the theory i.e. $R^{-1}$ and the second
$Z'$-boson associated with the extended gauge group i.e. $R^{-1}\leq
550$ GeV and $M_{Z'}\leq 1.2$ TeV and (ii) for this parameter range,
where the relic density of the KK neutrino contributes significantly
to the total dark matter relic density,
the DM-nucleon cross-section is $\geq 10^{-44}$ cm$^2$, a prediction
that is accessible to the next round of dark matter searches. No
signal in dark matter searches as well as the searches for the KK
modes and $Z'$ in the above range will rule out this class of model
and alternative solutions to the neutrino mass and proton decay
problem will have to be sought to keep the UED models
phenomenologically viable. Discovery of two components to dark
matter should also have implications for cosmology of structure
formation.

\section{The basic features of the model}
In this section, we review the basic features of the model in
Ref.\cite{abdel}. The gauge group of the model is $SU(3)_c\times
SU(2)_L\times SU(2)_R\times U(1)_{B-L}$ with matter content per
generation as follows:
\begin{eqnarray}
{\cal Q}_{1,-}, {\cal Q}'_{1,-}= (3,2,1,\tfrac{1}{3});&
{\cal Q}_{2,+}, {\cal Q}'_{2,+}= (3,1,2,\tfrac{1}{3});\nonumber\\
{\cal \psi}_{1,-}, {\cal \psi}'_{1,-}= (1,2,1,-1);& {\cal
\psi}_{2,+}, {\cal \psi}'_{2,+}= (1,1,2,-1); \label{matter}
\end{eqnarray}
where,  within parenthesis, we have written the quantum numbers that
correspond to each group factor, respectively and the subscript
gives the six dimensional chirality chosen to cancel gravitational
anomaly in six dimensions. Note that there are equal number of
positive and negative six dimensional chirality states. W e denote
the gauge bosons as $G_M$, $W^{\pm}_{1,M}$, $W^{\pm}_{2,M}$, and
$B_M$, for $SU(3)_c$, $SU(2)_L$, $SU(2)_R$ and $U(1)_{B-L}$
respectively, where $M=0,1,2,3,4,5$ denotes the six space-time
indices.  We will also use the following short hand notations: Greek
letters $\mu,\nu,\dots=0,1,2,3$ to denote  usual four dimensions
indices and lower case Latin letters $a,b,\dots=4,5$ for the extra
space dimensions. We will also use $\vec y$ to denote the
($x_4,x_5$) coordinates of a point in the extra space.

First, we compactify the extra $x_4$, $x_5$ dimensions into a torus,
$T^2$, with equal radii, $R$, by imposing  periodicity conditions,
$\varphi(x_4,x_5) = \varphi(x_4+ 2\pi R,x_5) = \varphi(x_4,x_5+ 2\pi
R)$ for any field $\varphi$.
 This has the effect of breaking the original $SO(1,5)$
Lorentz symmetry group of the six dimensional space  into the
subgroup $SO(1,3)\times Z_4$, where the last factor corresponds to
the group of discrete  rotations in the $x_4$-$x_5$ plane, by angles
of $k\pi/2$ for $k=0,1,2,3$. This is a subgroup of the continuous
$U(1)_{45}$ rotational symmetry contained in $SO(1,5)$. The
remaining $SO(1,3)$ symmetry  gives the usual 4D Lorentz invariance.
The presence of the surviving $Z_4$ symmetry leads to suppression of
proton decay\cite{yee} as well as neutrino mass\cite{abdel}.

Employing the further orbifolding conditions i.e. $ Z_2: \vec{y}
\rightarrow -\vec {y} \qquad \mbox{and}\qquad
 Z'_2: \vec{y}~' \rightarrow -\vec{y}~'$
for $\vec y = (x_4,x_5)$; and where $\vec{y}~' = \vec{y} - (\pi R
/2, \pi R/2)$, we can project out the zero modes and obtain the KK
modes by assigning appropriate $Z_2\times Z'_2$ quantum numbers to
the fields.

In the effective 4D theory the mass of each mode has the form:
$m_{N}^2 = m_0^2 + \frac{N}{R^2}$; with $N=\vec{n}^2=n_1^2 + n_2^2$
and $m_0$ is the Higgs vacuum expectation value (vev) contribution
to mass, and the physical mass of the zero mode.

We assign the following $Z_2\times Z'_2$ charges to the various
fields:
\begin{eqnarray}
G_\mu(+,+);\quad B_\mu(+,+);
\nonumber\\
W_{1,\mu}^{3,\pm}(+,+);W^3_{2,\mu}(+,+);
W^\pm_{2,\mu}(+,-); \nonumber\\
G_{a}(-,-);\quad B_a(-,-);\nonumber\\
\quad W_{1,a}^{3,\pm}(-,-); W^3_{2,a}(-,-); W^\pm_{2,a}(-,+).
\label{gparity}
\end{eqnarray}
As a result, the gauge symmetry $SU(3)_c\times SU(2)_L\times
SU(2)_R\times U(1)_{B-L}$
breaks down to $SU(3)_c\times SU(2)_L\times U(1)_{I_{3R}}\times
U(1)_{B-L}$ on the 3+1 dimensional brane. The $W^{\pm}_R$ pick up mass
$R^{-1}$ whereas prior to symmetry breaking the rest of the gauge bosons
remain massless.

For quarks we choose,
\begin{eqnarray}
 Q_{1,L}\equiv
   \left(\begin{array}{c} u_{1L}(+,+)\\ d_{1L}(+,+)\end{array}\right);
 &\quad&
 Q'_{1,L}\equiv
   \left(\begin{array}{c} u'_{1L}(+,-)\\ d'_{1L}(+,-)\end{array}\right);
   \nonumber \\
 Q_{1,R}\equiv
   \left(\begin{array}{c} u_{1R}(-,-)\\ d_{1R}(-,-)\end{array}\right);
 &\quad&
 Q'_{1,R}\equiv
   \left(\begin{array}{c} u'_{1R}(-,+)\\ d'_{1R}(-,+)\end{array}\right);
   \nonumber
\end{eqnarray}
\begin{eqnarray}
 Q_{2,L}\equiv
   \left(\begin{array}{c} u_{2L}(-,-)\\ d_{2L}(-,+)\end{array}\right);
 &\quad&
Q'_{2,L}\equiv
   \left(\begin{array}{c} u'_{2L}(-,+)\\ d'_{2L}(-,-)\end{array}\right);
   \nonumber \\
 Q_{2,R}\equiv
   \left(\begin{array}{c} u_{2R}(+,+)\\ d_{2R}(+,-)\end{array}\right);
& \quad& Q'_{2,R}\equiv
   \left(\begin{array}{c} u'_{2R}(+,-)\\ d'_{2R}(+,+)\end{array}\right);
\label{quarks} \end{eqnarray} and for leptons:
 \begin{eqnarray}
 \psi_{1,L}\equiv
   \left(\begin{array}{c} \nu_{1L}(+,+)  \\ e_{1L}(+,+)\end{array}\right);
  &\qquad&
 \psi'_{1,L}\equiv
 \left(\begin{array}{c} \nu'_{1L}(-,+)  \\ e'_{1L}(-,+)\end{array}\right);
   \nonumber \\
 \psi_{1,R}\equiv
    \left(\begin{array}{c} \nu_{1R}(-,-)  \\ e_{1R}(-,-)\end{array}\right);
  & \qquad &
 \psi'_{1,R}\equiv
  \left(\begin{array}{c} \nu'_{1R}(+,-) \\ e'_{1R}(+,-)\end{array}\right);
   \qquad
   \nonumber \\ [1ex]
 \psi_{2,L}\equiv
   \left(\begin{array}{c} \nu_{2L}(-,+) \\ e_{2L}(-,-)\end{array}\right);
  &\qquad&
 \psi'_{2,L}\equiv
    \left(\begin{array}{c} \nu'_{2L}(+,+)\\ e'_{2L}(+,-)\end{array}\right);
  \nonumber \\
 \psi_{2,R}\equiv
 \left(\begin{array}{c} \nu_{2R}(+,-)\\ e_{2R}(+,+)\end{array}\right);
&   \qquad &
 \psi'_{2,R}\equiv
  \left(\begin{array}{c} \nu'_{2R}(-,-)\\ e'_{2R}(-,+)\end{array}\right).
\label{leptons}
 \end{eqnarray}
 The zero modes i.e. (+,+) fields corresponds to the standard
 model fields along with an extra singlet neutrino which is
 left-handed. They will have zero mass prior to gauge symmetry
 breaking.

Turning now to the Higgs bosons, we choose a bidoublet, which will
be needed to give masses to fermions and break the standard model
symmetry
 and a pair of doublets $\chi_{L,R}$ with the
following $Z_2\times Z'_2$ quantum numbers:
\begin{align}
\phi &\equiv
\left(\begin{array}{cc} \phi^0_u(+,+) & \phi^+_d(+,-)\\
   \phi^-_u(+,+) &  \phi^0_d(+,-)\end{array}\right);\nonumber\\
\chi_L&\equiv \left(\begin{array}{c} \chi^0_L(-,+) \\
   \chi^-_L(-,+)\end{array}\right); \quad
\chi_R\equiv \left(\begin{array}{c} \chi^0_R(+,+) \\
   \chi^-_R(+,-)\end{array}\right),
 \end{align}
and the following charge assignment under the gauge group,
\begin{eqnarray}
\phi &=& (1,2,2,0),\nonumber\\
\chi_L&=&(1,2,1,-1),\quad \chi_R=(1,1,2,-1).
\end{eqnarray}
At the zero mode level, only the SM doublet $(\phi^0_u, \phi^-_u)$
and a singlet $\chi^0_R$ appear. The vev's of these fields, namely
$\langle\phi^0_u\rangle = v_{wk}$ and $\langle\chi^0_R\rangle= v_R$,
break the SM symmetry and the extra $U(1)_Y'$ gauge group,
respectively.

There are two classes of levels: one class corresponding to even KK
number with $Z_2\times Z'_2$ quantum numbers $(+,+)$ and $(-,-)$ and
another class corresponding to odd KK number, corresponding to
$Z_2\times Z'_2$ quantum numbers $(+,-)$ and $(-,+)$. Of these only
$(+,+)$ modes contain the zero mode as noted earlier. This implies
that the lightest KK modes are those in $(+,-)$ or $(-,+)$ class.

As there are a large number of KK modes, one may worry whether or not
electroweak precision constraints are satisfied.
To our knowledge,
there has been no such analysis for similar models, and it is outside
the scope of the current paper to perform a complete analysis regarding
the electroweak constraints.  Therefore, we leave the investigation of this
open issue for future work.

\section{Dark matter candidates}
In our UED model, there are the following stable KK modes: the
lowest KK excitation of the hypercharge boson, $B_Y^{KK}$ and the
lowest $(\pm,\mp)$ modes $\nu_{2,L}$ and $\nu_{2,R}$. Both the heavy
neutrino states couple only the $SU(2)_R$ gauge fields. The former
($B_Y^{KK}$) being a KK mode of the (+,+) state has twice the
mass-squared of the lowest KK modes of states of $(\pm,\mp)$ type
(i.e. $\nu_{2,L,R}$). The discussion of dark matter candidate has to
take this into account to see which particle really is the dark
matter. We find that in general it is an admixture of both. We also
include the effect of radiative corrections\cite{cheng1} which shift
the mass levels by an amount $\sim \frac{g^2n}{16\pi^2 R} \ln
(\tfrac{\Lambda}{\mu})$ where $n$ denotes the KK mode number,
$\Lambda$, the fundamental scale and $\mu$ the renormalization
point. When they are included, the values of the masses differ
slightly but the same lightest modes as identified here remain.

The $\nu^{\smbox{KK}}_{2R,L}$ couple to $Z'$; their annihilation
rate in the early universe will therefore be determined by $M_{Z'}$
which contributes in $s$-channel processes.  There are also
annihilation channels through $t$-channel processes mediated by
$W_2^{\pm}$, whose mass has a contribution from $R^{-1}$ as well as
$v_R$. The discussion of the annihilation channels of $B_Y^{KK}$ is
similar to that in \cite{tait,feng}.

\subsection{Annihilation Channels of $\nu_{2L,2R}$}
Since the yukawa couplings are small, annihilations through gauge
boson exchanges provide the dominant channel. Although we have two
independent Dirac fermions for dark matter, $\nu^{(10)}$ and
$\nu^{(01)}$, they couple the same way to $Z^{\prime}_{\mu}$ and
have the same annihilation channel. The only difference is that, for
charged current processes, $\nu^{(01)}$ ($\nu^{(10)}$) couples to
$W_{2,\mu}^{\pm,(01)}$ ($W_{2,\mu}^{\pm,(10))}$). The generic
coupling of matter fields to $Z_{\mu}^{\prime}$ is
\begin{eqnarray}
g(\overline{f}{f}Z^{\prime}_{\mu})\equiv
\tilde{g}_f=\frac{1}{2\sqrt{g_{BL}^2+g_R^2}}\left(-2T_3
g_R^2+Y_{BL}g_{BL}^2 \right),
\end{eqnarray}
where $T_3=\pm\tfrac{1}{2}$ for right-handed particles of Standard
Model and the sterile neutrino, and $T_3=0$ for left-handed
particles. We also have $Y_{BL}=+1/3$ for quarks and $Y_{BL}=-1$ for
leptons.  The cross section for
$\sigma(\overline{\nu}_2\nu_2\rightarrow \overline{f}f)$ from $Z'$
exchange can be written as
\begin{equation}
\sigma(\overline{\nu}_2\nu_2\rightarrow \overline{f}f)
v_{\smbox{rel}}= a_{Z^{\prime}}(f) +
b_{Z^{\prime}}(f)v_{\smbox{rel}}^2 \label{eq:S-general}
\end{equation}
where
\begin{eqnarray}
a_{Z^{\prime}}(f) &=&
\frac{\tilde{g}^2_{\nu}\tilde{g}^2_f}{2\pi}\frac{M^2}{(4M^2-M^2_{Z^{\prime}})^2}\nonumber \\
b_{Z^{\prime}}(f) &=&
\frac{\tilde{g}^2_{\nu}\tilde{g}^2_f}{2\pi}\frac{M^2}{(4M^2-M^2_{Z^{\prime}})^2}\left(\frac{1}{6}-\frac{2M^2}{4M^2-M^2_{Z^{\prime}}}\right).
\end{eqnarray}
\newline
\indent For the final state of $\overline{e}_R{e}_R$, we have a
$t$-channel process through charge current. The cross-section
therefore involves three pieces: two due to the squared-amplitudes
of $s-$ and $t-$channel diagrams and another from their
interference, denoted by $\sigma_{ss}$, $\sigma_{tt}$ and
$\sigma_{st}$ respectively.  Of these, $\sigma_{ss}$ has the same
form as Eq. \ref{eq:S-general}, and we will show that $\sigma_{tt}$
and $\sigma_{st}$ is parametrically smaller than $\sigma_{s}$.  Thus
the main contribution to the annihilation of $\nu_{2L,R}$ comes from
annihilation through $s$-channel processes mediated through
$Z^{\prime}_{\mu}$.

Due to $Z-Z^{\prime}$, there can also be annihilation of KK neutrino
into SM Higgs charged bosons.  In the limit that $v_w\ll v_R$,
we can work to the leading-order in the expansion of
$\mathcal{O}(v^2_w/v^2_R)$, where we can estimate these processes by treating
the $Z-Z^{\prime}$ mixing as a
mass-insertion.
In terms of Feynman diagrams, these annihilation channels are $s$-channel processes,
where a pair KK neutrino annihilates into a $Z^{\prime}$-boson, which propagates
to the mixing vertex, converting $Z^{\prime}$ to $Z$, which then decays into
$h^{\ast}h$ (both neutral and charged) or massless $W^+W^-$.
Compared to the
amplitude of annihilation of KK neutrino into SM fermions, the
annihilation to the bosons have effectively a replaced propagator
\begin{align}
\frac{1}{(s-M^2_{Z^{\prime}})}\rightarrow
\frac{1}{(s-M^2_{Z^{\prime}})} \delta\!M^2 \frac{1}{(s-M_Z^2)}
\end{align}
where
\begin{align}
\delta\!M^2\equiv\frac{\gR^2}{\sqrt{(\gL^2+\gY^2)(\gR^2+\gBL^2)}}M_Z^2,
\end{align}
is the off-diagonal element in the $Z-Z^{\prime}$ (mass)$^2$ matrix.
Since $s\sim 4M^2_{\nu}=4 R^{-2}$,  the annihilation cross section
into transverse gauge bosons and the Higgs bosons are suppressed by a factor of
$M_Z^4/s^2\sim (100 GeV)^4/ 16(500 GeV)^4\sim 10^{-4}$, and
can therefore be neglected.  As for the longitudinal modes, the ratio of annihilation cross-sections
of the longitudinal modes of the gauge bosons to the one single mode of SM fermion-antifermion pair is roughly
\begin{align}
\frac{\sigma(\nu^{KK}\nu^{KK}\rightarrow W^+W^-)}
{\sigma(\nu^{KK}\nu^{KK}\rightarrow \overline{f}{f})}
\sim  \left(\frac{\delta M^2}{m_W^2}\right)^2.
\end{align}
This ratio is about $\tfrac{1}{2}$ for $\gR=0.7 \gL$.  As there is only one annihilation mode into
the longitudinal modes of the charged gauge bosons, whereas there are many annihilation channels
to the SM fermion-antifermion pairs, the total annihilation cross section is dominated
by the SM fermion-antifermion contributions.

We note here that the annihilation
channels to matter fields differ from the analysis of \cite{tait}
and \cite{feng} in two important ways. First, in their
 analysis, the $s$-channel process is mediated by $Z$-boson of
the SM, whose mass can be ignored, whereas we have $s$-channel
processes mediated by $Z^{\prime}$, whose mass is significantly
higher than the mass of our dark matter candidate in the region of
interest.  Second, although we keep all contributions to the
annihilation cross sections for the KK neutrino $\nu_{2L,2R}$ in our
numerical work, to a workable approximation we can discard
$t$,$u$-channel processes mediated by charged gauge bosons
$W_2^{\pm}$, because $m^2_{W_2^{\pm}}$ has contributions both from
$R^{-1}$ and $v_R$. We have checked that excluding such processes
does not affect the main conclusions of the letter. To see this, let
us make the approximation $m^2_{W^{\pm}}=m^2_{Z^{\prime}}+R^{-2}$,
then we compare the cross section involving the product of a $t$ or
$u$ diagram with a $s$-channel diagram $\sigma_{st}$ with that
coming from the square of an $s$-channel diagram $\sigma_{ss}$,
\begin{align}
\frac{\sigma_{ss}}{\sigma_{st}}\approx
\frac{(s-m^2_{Z^{\prime}})(t-m^2_{W^{\pm}})}
{(s-m^2_{Z^{\prime}})^2}
=\frac{{2 (R^{-1})^2+m^2_{Z^{\prime}}}}{4
(R^{-1})^2-m^2_{Z^{\prime}}}.
\end{align}
Then $\sigma_{ss}\gg\sigma_{st}$ would require that
$m^2_{Z^{\prime}}> 2(R^{-1})^2$,
which is satisfied in the region of interest in the parameter space.
Similarly, the cross section involving two $t-$ or $u$-channel
diagrams, $\sigma_{tt},\sigma_{uu}$ or $\sigma_{tu}$ is small
compared to $\sigma_{ss}$.
Thus the annihilation cross section is given by
\begin{align}
\sigma(\overline{\nu}_2\nu_2\rightarrow XX)v_{\smbox{rel}} \simeq
\sum_{{SM}}(a_{Z^{\prime}}(f) + b_{Z^{\prime}}(f)v_{\smbox{rel}}^2),
\end{align}

\subsection{Relic Density: $\nu^{KK}_{2L,2R}$ vs $B_Y^{KK}$}
\begin{figure}[tbp]
\begin{center}
\includegraphics[width=2.7in]{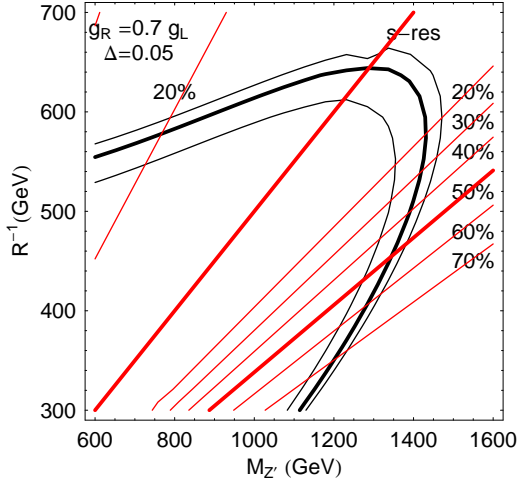}
\caption{The contour in the $1/R-M_{Z'}$ which corresponds to
$\Omega_{\nu_{L,R}}h^2 + \Omega_{B_Y}h^2$  being the observed dark
matter. The intersection of the red lines with the contour indicate
the the fraction of KK neutrinos in the dark matter.}
\end{center}
\end{figure}
As noted, the contribution of $B_Y^{KK}$ to relic density in our
case does not differ from the calculation of \cite{tait,feng} since
our model in the low energy limit all the $B_Y^{KK}$ annihilation
modes are same as the model used there. Combining the RH neutrino
and $B_Y^{KK}$ contributions, we get the total relic density.
Clearly, for different regions of parameter space, the relative
fraction of the two components will be different. In Fig. 1, we
combine the two contributions and plot the allowed regions in the
$R^{-1}$ and $M_{Z'}$ parameter space so that we get the
$\Omega_{\nu_{L,R}}h^2 + \Omega_{\gamma^{(1)}}h^2$ to the observed
value\cite{wmap}. The central solid line corresponds to the central
value of the dark matter contribution to $\Omega h^2$ and the outer
lines denote the one $\sigma$ error range. The straight lines in
Fig.1 give the parameter range of $M_{Z'}$ and $R^{-1}$ for which we
get the noted fraction of the RH neutrino contribution to $\Omega
h^2$. For small values of $M_{Z^{\prime}}$, the annihilation of
$\nu^{\smbox{(01)}}_{2L,2R}$ is efficient and most of the dark
matter is $B_Y^{KK}$ having a mass of roughly $\sqrt{2}R^{-1}\sim
700$ GeV.  In fact, along the line
$2M_{\nu^{(01)}}=2R^{-1}=M_{Z^{\prime}}$, the annihilation of
$\nu^{\smbox{(01)}}_{2L,2R}$ has an $s$-channel resonance, and its
contribution to dark matter relic density is minimal. Away from the
line of $s$-channel resonance, the contribution of
$\nu^{\smbox{(01)}}_{2L,2R}$ to the relic density increases, and
$R^{-1}$ decreases so as to decrease the relic density due to
$B_Y^{KK}$, keeping the total relic density within the allowed
range. Using the present bounds on the $M_{Z'}$ of the left-right
model from collider data of $M_{Z'}\geq 860$ GeV\cite{pdg}, we
conclude that in our picture we have an interesting region in the
parameter space where the KK sterile neutrinos constitutes at least
30\% of the dark matter density when $M_{Z'}\gtrsim 1.2$ TeV and
$400<R^{-1}<550$ GeV.

\begin{figure}[htb]
\begin{center}
\includegraphics[width=3.3in]{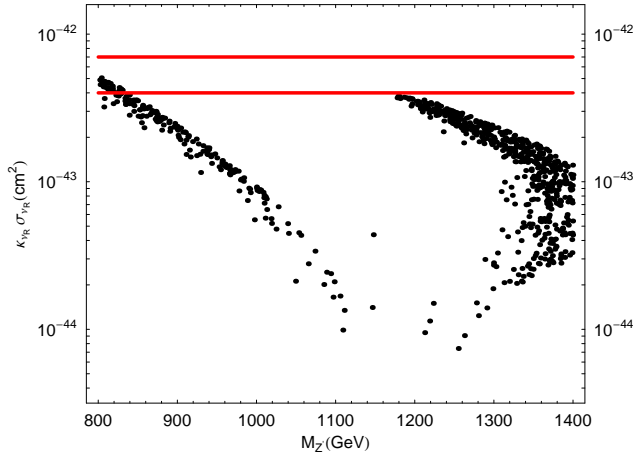}
\caption{Scattering  plot of the scattering cross section of
$\nu^{(1)}$ on the nucleon as function of $M_{Z'}$.
$\kappa_{\nu_R}\in \{0 - 0.7\}$ is the fractional contribution of KK
neutrino to the dark matter relic density.  (The value $\kappa =0$
corresponds to regions of $s$-channel resonance for $\nu^{KK}$
annihilation.)  The upper (lower) horizontal line corresponds to the
CDMS II upper bound for a dark matter of mass $500\;$ GeV ( $300\;$
GeV).}
\end{center}
\end{figure}

\subsection{Direct Detection of $\nu_{2L,2R}$}
As we have a two-component dark matter, the total dark
matter-nucleon cross section is given by
\begin{align}
\sigma_n=\kappa_{\nu_R}\sigma_{\nu_R}+\kappa_B \sigma_B,
\end{align}
where $\sigma_{\nu_R(B)}$ is the spin-independent KK neutrino
(hypercharge boson)- nucleon scattering  cross section  , and
\begin{align}
\kappa_{\nu_R}\equiv \frac{\Omega_{\nu_R} h^2 }{\Omega_{\nu_R} h^2
+\Omega_{B_Y^{KK}} h^2 },
\end{align}
is the fractional contribution of the KK neutrino relic density
to the total relic density of the dark matter.  $\kappa_B$ is
similarly defined.  As pointed out in Ref. \cite{tait},
$\sigma_B$ is of the order $\sigma_B\sim10^{-10}$ pb, and we will find that
$\sigma_{\nu_R}\gg\sigma_B$.  Therefore, it is a good approximation to
take $\sigma_n$ as
\begin{align}
\sigma_n\approx\kappa_{\nu_R}\sigma_{\nu_R}.
\end{align}

The $\nu_{2L,2R}$ scattering cross section per nucleon in a nucleus 
$N(A,Z)$ is given by
\begin{align}
\sigma_{\nu_R}&=\frac{b_N^2 m^2_n}{\pi A^2},
\end{align}
where $b_N=Z b_p+(A-Z)b_n$, and $b_{p,n}$ is the effective
four-fermion coupling between $\nu_{2L,2R}$ and a proton or neutron. They are
given by $b_p=2b_u+b_d$ and $b_n=b_u+2b_d$, with
\begin{align}
b_q&=\frac{g_{(\overline{\nu_2}{\nu_2}Z^{\prime})}}{2M^2_{Z^{\prime}}}
\sum_{i=L,R} \left[g_{(\overline{q}_i{q}_iZ^{\prime})}
-g_{(\overline{q}_i{q}_iZ)} \frac{\delta\!M^2}{M^2_Z}
+\mathcal{O}\left(\frac{M^2_Z}{M^2_{Z^{\prime}}}\right) \right],
\end{align}
so that we have taken into account the $Z-Z^{\prime}$ mixing up to
$\mathcal{O}(v^2_w/v^2_R)$.
The contribution due to $Z-Z^\prime$ mixing can be understood 
diagrammatically as
a $Z^{\prime}$ propagator of $M_{Z^{\prime}}^{-2}$ followed by a
mass-insertion of $\delta\!M^2$, followed by a $Z$-propagator of
$M_Z^{-2}$.  This is equivalent to a mixing term of
$\delta\!M^2/M_{Z^{\prime}}^{2}$ multiplying a $Z$-propagator of
$M_Z^{-2}$ at leading order in $M^2_Z/M^2_{Z^{\prime}}$.

For $M_{Z^{\prime}}=1.2$ TeV, $g_R=0.7 g_L$, $A=73$ and $Z=32$ (for
the Ge detectors used at CDMS II), we obtain $\sigma_n=3.87\times
10^{-43}\mbox{cm}^2$. In Fig. 2, we give the scatter plot of the
predicted values of the scattering cross section between $\nu^{(1)}$ and
the nucleon as function of $M_{Z'}$ for $\kappa_{\nu_R}\in \{0 -
0.7\}$.
The horizontal lines correspond to the upper bounds on
$\sigma_n$ from CDMS II
for dark matter candidates with masses 300
and 500 GeV, which are about 4$\times
10^{-43} \mbox{cm}^2$ and 7$\times
10^{-43} \mbox{cm}^2$, respectively.
We find that if this cross section is probed down to the level of
$10^{-8}$ picobarns, the regions with large $\kappa_{\nu_R}$
(corresponding to $M_{Z^{\prime}}\sim 1300$ GeV and $400
\mbox{GeV}<R^{-1}<500 \mbox{GeV}$) can be tested. We believe that
this makes this model quite interesting.

\section{Conclusions}
In summary, we have proposed a new dark matter scenario which
consists of an admixture of two Kaluza-Klein modes: the right handed
neutrinos of a left-right symmetric model embedded into a
six-dimensional brane-bulk theory i.e. $\nu^{\smbox{KK}}_{2L,R}$ and
$B^{KK}_Y$. This class of universal extra dimensional models solve
naturally both the proton stability and neutrino mass problem by an
extended electro-weak gauge group combined with appropriate orbifold
quantum numbers for fermions. Our detailed analysis of dark matter
constraints i.e. $\Omega h^2$ and DM-nucleon cross-section leads us
to predict the existence of an extra $Z'$ boson with mass less than
1.4 TeV which should be accessible at the LHC.
We also find that, for the region of parameter space where the relic abundance of $\nu^{\smbox{KK}}_{2L,R}$
contributes significantly to the total relic density of dark matter,
the DM-nucleon cross-sections are above $10^{-8}$ picobarns,
which can be
explored in the ongoing and planned dark matter
search experiments \cite{cdms,crest}.

 This work is supported by the National Science Foundation grant
no. Phy-0354401 and the UMD Center for Particle and String Theory.

\end{document}